\documentclass{xws-ijmpa}

\newcommand{\be}{\begin{equation}}
\newcommand{\ee}{\end{equation}}
\newcommand{\ba}{\begin{eqnarray}}
\newcommand{\ea}{\end{eqnarray}}
\newcommand{\se}{\setcounter{equation}{0}}
\newcommand{\re}[1]{(\ref{#1})}
\renewcommand{\theequation}{\arabic{section}.\arabic{equation}}

\newcommand{\1}{^{-1}}
\newcommand{\B}{{\cal B}} 
\newcommand{\dg}{^{\dagger}}
\newcommand{\di}{\mbox{d}\,}
\newcommand{\e}{\mbox{e}}
\newcommand{\f}{{\mbox{\scriptsize f}}} 
\newcommand{\g}{{\mbox{\scriptsize g}}} 
\newcommand{\G}{{\cal G}} 
\newcommand{\ga}{\gamma_5}
\newcommand{\h}{\frac{1}{2}}
\newcommand{\Hw}{{\cal H}} 
\newcommand{\Id}{\mbox{1\hspace{-.9mm}l}}   
\newcommand{\ka}{\kappa}
\newcommand{\la}{\lambda}
\newcommand{\mi}[1]{_{\mbox{\tiny #1}}}
\newcommand{\mii}[2]{_{\mbox{\tiny #1}\,#2}}
\newcommand{\mo}[1]{^{(\mbox{\scriptsize #1})}}
\newcommand{\na}[1]{\nabla_{#1}}
\newcommand{\Nh}{\hat{N}}
\newcommand{\ra}{\rightarrow}
\newcommand{\sy}{\scriptscriptstyle} 
 
\newcommand{\Tr}{\mbox{Tr}} 
 
\newcommand{\vp}{\varphi} 
\newcommand{\W}{_{\mbox{\tiny W}}}

\begin{document}

\markboth{W.~Kerler}
{Chiral Fermion Operators on the Lattice}

\catchline{}{}{}

\title{CHIRAL FERMION OPERATORS ON THE LATTICE}

\author{\footnotesize WERNER KERLER}

\address{Institut f\"ur Physik, Humboldt-Universit\"at, D-10115 Berlin,
Germany\\E-mail: kerler@physik.hu-berlin.de}

\maketitle

%\pub{Received (Day Month Year)}{Revised (Day Month Year)}

\begin{abstract}
We only require generalized chiral symmetry and $\ga$-hermiticity, which
leads to a large class of Dirac operators describing massless fermions
on the lattice, and use this framework to give an overview of developments
in this field. Spectral representations turn out to be a powerful tool for
obtaining detailed properties of the operators and a general construction 
of them. A basic unitary operator is seen to play a central r\^ole in this 
context. We discuss a number of special cases of the operators and elaborate
on various aspects of index relations. We also show that our weaker 
conditions lead still properly to Weyl fermions and to chiral gauge theories.

%\keywords{Keyword1; keyword2; keyword3.}
\end{abstract}

\section{Introduction}\label{INT}

In investigations of chiral fermions on the lattice the Dirac operator $D$ is
largely required to satisfy the Ginsparg-Wilson (GW) relation\cite{gi82}
\be
\{\ga,D\}=\rho^{-1}D\ga D\,,
\label{GW}
\ee
with a real constant $\rho$, and to be $\ga$-hermitian,
\be
D\dg=\ga D\ga\,. 
\label{ga5a}
\ee
In this GW case L\"uscher has pointed out that the classical action has a
generalized chiral symmetry which corresponds to the condition\cite{lu98} 
\be
\ga D+D\hat{\gamma}_5=0\quad\mbox{ with }\quad\hat{\gamma}_5=\ga(\Id-\rho\1 D)
\label{hga}
\ee
for the Dirac operator $D\,$.

We note that conditions \re{GW} and \re{ga5a} imply that the operator 
$\Id-\rho\1 D$ is unitary and $\ga$-hermitian. Therefore requiring $D$ to 
have the form
\be
D=\rho(\Id-V) \quad\mbox{ with }\quad V\dg=V^{-1}=\ga V \ga
\label{DN}
\ee
is equivalent to imposing \re{GW} and \re{ga5a}. Further this indicates that
\re{hga} actually has the general form 
\be
\ga D + D \ga V = 0 \quad\mbox{ with }\quad V\dg=V^{-1}=\ga V \ga\,.
\label{gga}
\ee

We have recently found\cite{ke02a} that, instead of imposing \re{GW} and 
\re{ga5a} as is usually done, only requiring \re{gga} and \re{ga5a} leads to
a large class of Dirac operators describing massless fermions on the lattice,
which includes GW fermions, the ones proposed by Fujikawa\cite{fu00} and the 
extension of the latter\cite{fu02} as special cases. Nevertheless it has 
turned out that our weaker conditions still lead properly to Weyl fermions 
and to chiral gauge theories. 

On the basis of our conditions we have performed a detailed analysis of the 
properties of the Dirac operators as well as worked out a general construction
of them. This has become possible because for the general class $D$ gets a 
function of $V$, so that the spectral representation of $V$ can be used to 
deal with $D=F(V)\,$. In addition to the conditions \re{gga} and \re{ga5a}, 
we have required that $F$ must allow for a nonvanishing index of the Dirac 
operator.

In the present paper we use this framework to give an overview of developments
in the field of massless fermions on the lattice. We start with general
properties of the Dirac operators and index relations. We then turn to the
general construction of such operators and after this to the discussion of
special cases. Next various issues related to the index are addressed.
Finally we consider chiral gauge theories.

In Sec.~\ref{PROPD} we first give the basic relations which immediately
follow from our general conditions. We then introduce spectral representations
and obtain related properties. After this we derive various relations for 
the index of $D$ in a general way. It turns out that solely the operator $V$ 
enters the index for the general class of $D\,$. The result generalizes ones 
obtained in the overlap formalism\cite{na93} and in the GW 
case\cite{ha98,lu98i} before. Similarly the sum rule, first observed by 
Chiu\cite{ch98} in the GW case, is found to hold in general and to rely 
only on $V\,$.

In Sec.~\ref{CONS} we describe the general construction of Dirac operators of
the class. We use spectral functions $f$, which on the on hand side describe 
the location of the spectum of $D\,$, and on the other provide a powerful tool
for the construction. The general form of $f$ is found to involve two 
functions, $w$ and $h\,$, which must have specific properties and determine
the form of $D\,$. Particular choices of $h$ turn out to require related forms
of $V\,$. We give a realization of such $V$ which generalizes the $V$ 
implicit in the overlap Dirac operator of Neuberger.\cite{ne98}

In Sec.~\ref{SPECA} we discuss the special cases which can be found 
literature. We start with the GW relation, also adding some remarks about 
properties of its general form. We then show that the extended Fujikaw 
type\cite{fu02} belongs to the general class and bring $D$ of the respective
original proposal\cite{fu00} into the form $D=F(V)\,$. 

In Sec.~\ref{SPECB} we study further special cases. First, to see effects 
of nontrivial (nonconstant) $w$ we use trivial $h(x)=x\,$. In a special case 
of this $D$ is given by an expansion in powers of $V$. In examples of 
nontrivial $w$ together with nontrivial $h(x)$ it becomes still explicit how 
conditions on $w$ are related to behaviors at the eigenvalues $\pm1$ of $V\,$. 
We then present a method for getting a subclass of nontrivial functions 
$h(x)\,$ and work out a particular example of this.

In Sec.~\ref{FLOW} we consider the relation between the index of the Dirac
operator $D$ and the spectral flows of the hermitian Wilson-Dirac operator. 
This relation, which is important in the overlap formalism,\cite{na93} turns 
out to extend to the generalized  Wilson-Dirac operator considered here. We 
also derive the differential equation\cite{ke99} which describes these 
spectral flows in an exact way.

In Sec.~\ref{SUML} we address questions which arise for the index relations 
in the continuum limit. After an overview of the relevant issues, we make 
sure that the sum rule for the index still holds in the limit. For this 
purpose we firstly give the proper definition of the trace expressions 
involved and secondly show that a continuous part of the spectrum does 
not contribute. We then elaborate on the structural difference to the 
Atiyah-Singer framework\cite{at68} which is reflected by the sum rule.

In Sect.~\ref{CAY} we turn to the alternative form of $V$ realized in work
by Chiu\cite{ch01} and found not to describe topological properties on the
finite lattice correctly. This form is of the Cayley-transform type for
which we have shown\cite{ke02} that correct properties generally only arise
in the limit. After some remarks about explicit $V$ we present the 
analysis of Cayley-type $V$. We then also make details of the spaces involved 
precise.

In Sec.~\ref{CHIR} we show that the general class of operators considered
leads still properly to Weyl fermions and to chiral gauge theories. In this
context it turns out that quantities like the chiral degrees of freedom 
or the gauge anomaly only involve $V$ for the whole class. We also add
some remarks about proper accounting for zero modes of $D\,$.

\section{Properties of Dirac Operators}\se\label{PROPD}

\subsection{Basic Relations}

As pointed out in the introduction, we impose the conditions
\be
\ga D + D \ga V = 0 \quad\mbox{ with }\quad V\dg=V^{-1}=\ga V \ga\,,
\label{gg}
\ee
\be
D\dg=\ga D\ga 
\label{ga5}
\ee
on $D\,$. We first note that using \re{ga5} one gets from \re{gg} 
\be
D+D\dg V=0 \,,\quad D\dg+DV\dg=0\,,
\label{gg1}
\ee
by which it follows that $V$ and $D$ commute and that $D$ is normal, 
\be
[V,D]=0\,,\quad
DD\dg=D\dg D\,.
\label{COM}
\ee
To account for this we require $D$ to be a function of $V\,$,
\be
D=F(V)\,,
\ee 
i.e.~to depend on $V$ and possibly on constants, however, not on any other 
operator.

\subsection{Spectral Representations}

On the finite lattice $V$ has the spectral representation $V=\sum_kv_kP_k$ with
eigenvalues satisfying $|v_k|=1$ and orthogonal projections $P_k=P_k\dg\,$.
Therefore the operator functions $F(V)$ can be represented by 
$F(V)=\sum_kf(v_k)P_k\,$ with spectral functions $f(v)\,$. 
The use of this will be an important tool in the following.

Because of the $\ga$-hermiticity of $V$ its part related to real eigenvalues
commutes with $\ga$ and its complex eigenvalues come in pairs. Taking 
$V\dg=\ga V\ga$ into account we obtain the more detailed representation
\be
V=P_1^{(+)}+P_1^{(-)}-P_2^{(+)}-P_2^{(-)}+\sum_{k\;(0<\vp_k<\pi)}
\big(\e^{i\vp_k} P_k\mo{I}+\e^{-i\vp_k} P_k\mo{II}\big)\,,
\label{specv}
\ee
in which the projections satisfy 
\be
\ga P_j^{(\pm)}=P_j^{(\pm)}\ga= \pm P_j^{(\pm)}\,,\quad \ga P_k\mo{I}=
P_k\mo{II}\ga\,. 
\label{PPg}
\ee
The spectral representation of $D=F(V)$ then becomes 
\ba
D=f(1)\big(P_1^{(+)}+P_1^{(-)}\big)+f(-1)\big(P_2^{(+)}+P_2^{(-)}\big)\nonumber
\\ +\sum_{k\;(0<\vp_k<\pi)}\Big(f(\e^{i\vp_k})P_k\mo{I}+f(\e^{-i\vp_k})
P_k\mo{II}\Big) \,,
\label{specd}
\ea
in which $D$ is characterized by the function $f(\e^{i\vp})$. Clearly this
function enters only at the values $\e^{i\vp}=\e^{i\vp_k}\,$. However, 
since we want to define $D$ in general (in particular, for any gauge field 
configuration), we have to specify $f(v)$ for all $v=\e^{i\vp}\,$.

Inserting the general form \re{specd} into \re{ga5} and \re{gg1} we obtain
the conditions 
\be
f^*(v)=f(v^*)\,,
\label{cond*}
\ee
\be
f(v)+f^*(v)\,v=0\,,
\label{cond+}
\ee
respectively, for the functions $f(v)\,$. These conditions imply 
\be
f(1)=0\,,\qquad f(-1) \;{\rm real}\,.
\ee
The operator form of \re{cond*} is
\be
F\dg(V)=F(V\dg)\,.
\label{cond*o}
\ee

In addition to giving $D$ by \re{specd}, the functions $f(v)$ obviously 
describe the location of its spectrum. For continuous $f(\e^{i\vp})$ the 
eigenvalues of $D$ reside on a closed curve in the complex plane which is 
symmetric to the real axis and meets this axis at zero and at the value 
$f(-1)$ (which is required to be nonzero below).

\subsection{Index Relations}

Denoting the dimensions of the right-handed and of the left-handed eigenspace 
for eigenvalue $\pm1$ of $V$ by $N_{+}(\pm1)$ and $N_{-}(\pm1)$, 
respectively, we have from \re{PPg}
\ba
\mbox{Tr}\big(\ga P_1^{(\pm)}\big)=\pm N_{\pm}(1)\,,\quad
\mbox{Tr}\big(\ga P_2^{(\pm)}\big)=\pm N_{\pm}(-1)\,,\nonumber\\
\mbox{Tr}\big(\ga P_k\mo{I}\big)=\mbox{Tr}\big(\ga P_k\mo{II}\big)=0\,.
\hspace*{33.2mm}
\label{TP}
\ea
With this because of $f(1)=0\,$, using the resolvent $(D-\zeta\Id)^{-1}\,$, 
we obtain for the index of $D$ 
\be 
\lim_{\zeta\rightarrow 0}\mbox{Tr}\Big(\ga\frac{-\zeta}{D-\zeta\Id} \Big)= 
\left\{\begin{array}{ll} N_+(1)-N_-(1) &\mbox{for }f(-1)\ne0\\
N_+(1)-N_-(1)+N_+(-1)-N_-(-1) &\mbox{for }f(-1)=0 \end{array}\right.
\label{IND}
\ee
and also find
\be 
\lim_{\zeta\rightarrow 0}\mbox{Tr}\Big(\ga\frac{D}{D-\zeta\Id} \Big)=
\left\{\begin{array}{ll} N_+(-1)-N_-(-1) &\mbox{for }f(-1)\ne0\\
0 &\mbox{for }f(-1)=0 \end{array}\right. \,.
\label{IND1}
\ee
Adding up \re{IND} and \re{IND1} the sum on the l.h.s.~gets 
$\mbox{Tr}(\ga\Id)=0$ so that in any case
\be  
N_+(1)-N_-(1)+N_+(-1)-N_-(-1)=0\,.
\label{sum}
\ee
Because of \re{IND} and \re{sum}, to admit a nonvanishing index we have to 
impose the condition 
\be
f(-1)\ne0\,.
\label{conda}
\ee

After having \re{conda}, according to \re{sum} to allow for a nonvanishing 
index one has also to require that in addition to $1$ the eigenvalue $-1$ of 
$V$ occurs. The sum rule \re{sum} corresponds to the one found in 
Ref.~\refcite{ch98} for the special case of Dirac operators which satisfy 
the GW relation \re{GW}.

Using \re{specv} with \re{TP} we find
\be
\mbox{Tr}(\ga V)=N_{+}(1)-N_{-}(1)-N_{+}(-1)+N_{-}(-1)\,,
\ee
so that with \re{sum} we generally get for the index of the Dirac operators $D$
\be
N_+(1)-N_-(1)=\h\mbox{Tr}(\ga V)\,.
\label{INN}
\ee
Thus it turns out that solely the operator $V$ enters for the whole class.
This generalizes the results obtained in the overlap formalism\cite{na93}
and in the GW case\cite{ha98,lu98i} before.

In this context it is to be noted that the spectral representation of $\ga V$ 
on the basis of \re{specv} becomes
\be
\ga V= P_1^{(+)}-P_1^{(-)}-P_2^{(+)}+P_2^{(-)}+
              \sum_k\big(\check{P}_k^{[+]}-\check{P}_k^{[-]}\big)\,,
\label{dV}
\ee
where the projections $\check{P}_k^{[\pm]}$ are expressed in terms of 
$P_k\mo{I}$ and $P_k\mo{II}$ by
\be
\check{P}_k^{[\pm]}=\h\big(P_k\mo{I}+P_k\mo{II}\pm\e^{i\vp_k}\ga P_k\mo{I}
\pm\e^{-i\vp_k}\ga P_k\mo{II}\big)\,,\qquad 0<\vp_k<\pi\,.
\ee
With \re{TP} one gets 
\be
\mbox{Tr}\,\check{P}_k^{[+]}
=\mbox{Tr}\,\check{P}_k^{[-]}=\mbox{Tr}\,P_k\mo{I}=\mbox{Tr}\,P_k\mo{II}\,
\ee
for the respective dimensions.

\section{Construction of Dirac Operators}\se\label{CONS}

\subsection{Construction with Spectral Functions}

To develop a general construction of the Dirac operators of the class we
use the functions $f$ of the spectral representation \re{specd} as a tool. 
We start noting that condition \re{cond+}, $f(\e^{i\vp})+f^*(\e^{i\vp})
\e^{i\vp}=0\,$, can be written as
\be
\big(i\e^{-i\vp/2}f(\e^{i\vp})\big)^*=i\e^{-i\vp/2}f(\e^{i\vp})\,.
\label{SY1}
\ee
This shows that $f$ is of form
\be
f(\e^{i\vp})=-i\e^{i\vp/2}\,g(\vp)\,,\quad g(\vp)\;\mbox{real}\,.
\label{SY2}
\ee
Then from condition \re{cond*}, $f^*(\e^{i\vp})=f(\e^{-i\vp})$, the 
requirement
\be 
g(-\vp)=-g(\vp)
\label{SY3}
\ee
follows. Further, with $(2\pi)$-periodicity in $\vp$ of $f(\e^{i\vp})\,$, 
the function $g(\vp)$ has to satisfy
\be
\quad g(\vp+2\pi)=-g(\vp)\,.
\label{SY4}
\ee
We now see that forms of $D=F(V)$ can be obtained by determining functions 
$g(\vp)$ which are real, odd, and satisfy \re{SY4}.

In view of the indicated requirements the basic building blocks for the 
construction of $g$ are the functions $\,\sin\nu\vp/2$ and 
$\,\cos\mu\vp/2$ with integer $\nu$ and $\mu\,$. Using them one
arrives at the form  
\be
g=\sum_{\nu}s_{\nu}w_{\nu}(t_1,t_2,\ldots)\,, 
\label{GE1a}
\ee 
where $w_{\nu}$ are real functions and 
\be
s_{\nu}=\sin(2\nu+1)\vp/2\,,\quad t_{\mu}=\cos\mu\vp\,,\qquad\nu,\mu
\;{\rm integer}\,.
\label{GE1}
\ee
Because of the identity
\be
s_{\nu}=s_0 \Big(1+2\sum_{\mu=1}^{\nu}t_{\mu}\Big)
\label{GE2}
\ee
\re{GE1a} can be simplified to 
\be
g=s_0\,w(t_1,t_2,\ldots)
\label{GE2a}
\ee 
with a real function $w\,$. Further, since the $t_{\nu}$ are given
by polynomials of $t_1\,$,
\ba
t_{2\mu}=d_{2\mu}t_1^{2\mu}+d_{2\mu-2}t_1^{2\mu-2}+\ldots+d_0\,,
\hspace*{13mm} \nonumber\\
t_{2\mu+1}=d_{2\mu+1}t_1^{2\mu+1}+d_{2\mu-1}t_1^{2\mu-1}+\ldots+d_1t_1\,,
\label{GE3}
\ea
\re{GE2a} can be cast into the still simpler form 
\be
g=s_0\,w(t_1)\,.
\label{GE4}
\ee

We next note that given a function $g$ with the required properties, then 
$h(g)$ is again a function with such properties provided that $h$ is odd 
and real, 
\be
h(-x)=-h(x)\,,\quad h^*(x)=h(x)\;\;\mbox{for real}\;\;x\,.
\label{GE5}
\ee
With this the form \re{GE4} generalizes to
\be
g=h\big(s_0\,w(t_1)\big)\,.
\label{GE6}
\ee

To satisfy condition \re{conda}, $f(-1)\ne0\,$, we need $g(\pi)\ne0$ or 
$h\big(w(-1)\big)\ne0\,$. Therefore we have to impose
\be
w(-1)\ne0\,,
\label{W-1}
\ee
which is sufficient if $h(x)$ gets only zero for $x=0\,$. To guarantee this 
we in addition require strict monotony,
\be
h(x_2)>h(x_1)\;\;{\rm for}\;\;x_2>x_1\,.
\label{MON}
\ee
Then also the inverse function $\eta(y)$ with 
\be
\eta\big(h(x)\big)=x
\label{ETA}
\ee
is uniquely defined and strictly monotonous, which we will need below.

With \re{SY2}, \re{GE6} and \re{GE1} we now have the general form 
\be
f(\e^{i\vp})=-i\e^{i\vp/2}\,h\big(s_0\;w(t_1)\big)
=-i\e^{i\vp/2}\,h\big(\sin\!\frac{\vp}{2}\;w(\cos\vp)\big)
\label{GAi}
\ee
of the spectral function $f\,$.

\subsection{Nontrivial Choices of $h(x)$}

Looking for functions $h$ one has to note that because of the identity
\be
s_0^{2k+1}=s_0\,\sum_{\nu=0}^k\Big(\begin{array}{c}\scriptstyle 2k+1\\
\scriptstyle\nu\\\end{array}\Big) (-1)^{\nu}\Big(1+2\sum_{\mu=1}^{k-\nu}
t_{\nu}\Big)\,,\quad k=0,1,2,\ldots\,,
\label{GE7}
\ee 
\re{GE6} reduces to the form \re{GE4} if $h(x)$ is a polynomial or allows an 
expansion in powers of $x$. This limits the possibilities for nontrivial 
choices of $h(x)$, i.e.~of ones by which \re{GE6} gives something beyond 
\re{GE4}.

The nontrivial choices of $h$ actually are equivalence classes, i.e.~equivalent
ones are not to be counted as different. For example, $h(x)=x^{1/(2k+1)}$ and
$h(x)=x^{1/(2k+1)}r(x)$ with $r(-x)=r(x)$ are equivalent, because with $r(x)=
r(|x|)$ and $|s_0|=\sqrt{\h(1-t_1)}$ they give the same form in \re{GE6}. 
Also forming in addition odd powers because of \re{GE7} gives nothing new.

To get a criterion for triviality of $h$ we note that, given 
$g=h(s_0w(t_1))\,$, in the trivial case $\hat{w}(t_1)$ should exist so
that this could also be expressed as $g=s_0\hat{w}(t_1)\,$. For $\vp\ne0$ 
one gets simply $\hat{w}(t_1)=h(s_0w(t_1))/s_0\,$, while at $\vp=0$ 
conditions on $h$ and $w$ arise from 
\be
\lim_{s_0\ra0}|h\big(s_0w(t_1)\big)/s_0|<\infty\,,
\label{NTR}
\ee
which has to hold in the trivial case.

\subsection{Operator Form of Construction}

The operator $D$ for the general construction is obtained inserting \re{GAi}
into \re{specd}, which gives 
\be
D=-iV^{\h}\,H\Big(\frac{1}{2i}(V^{\h}-V^{-\h})\;W\big(\textstyle{\h}(V+V\dg)
\big)\Big)\,,
\label{GAO}
\ee
where $V^{\h}$ is defined with $+\e^{i\vp/2}$ in its spectral representation
corresponding to $\e^{i\vp}$ in that of $V\,$.
According to the reality of $w$ and to \re{W-1} $W$ has to satisfy
\be
W\dg(X)=W(X\dg)\,,\quad W(-P)\ne0\;\;{\rm for}\;\;P=P^2=P\dg>0\,.
\label{GA1i}
\ee
Because of \re{GE5} and \re{MON} for the function $H$ one needs
\ba
H(-X)=-H(X)\,,\qquad H\dg(X)=H(X)\hspace*{25mm}\nonumber\\
H(X)>H(X_1)\;\;{\rm for}\;\;X>X_1\quad\mbox{ with } X\dg=X\,,X_1\dg=X_1\,.
\label{MONO}
\ea
Then corresponding to \re{ETA} also the inverse function $E$ of $H$ 
\be
E\big(H(X)\big)=X 
\ee
exists and is odd, hermitian and strictly monotonous, too, i.e.~satisfies
analogous relations as for $H$ are given in \re{MONO}. Equation \re{GAO} 
demonstrates the drastic increase of possible forms of Dirac operators 
which occurs as compared to the GW case \re{GW}.

\subsection{Realization of $V$}\label{REA}

To specify $V$ explicitly we introduce the normalization-type definition
\be
V=-D\W^{(\eta)}\Big(\sqrt{D\W^{(\eta)\dag}D\W^{(\eta)}}\,\Big)\1
\label{Vg}
\ee
with
\be
D\W^{(\eta)}=iE\Big(\frac{1}{2i}\sum_{\mu}\gamma_{\mu}(\na{\mu}-\na{\mu}\dg)
\Big)+E\Big(\frac{r}{2}\sum_{\mu}\na{\mu}\dg\na{\mu}\Big)+E\big(m\Id\big) \,.
\label{VgD}
\ee
Obviously for $E(X)=X$ \re{VgD} reduces to the overlap form of 
Neuberger.\cite{ne98}

We note that the strict use of hermitian functions of hermitian operators in 
\re{Vg} with \re{VgD} makes it generally applicable. Then monotony is also 
defined in terms of operators. In particular, it will allow below to evaluate 
the limit in a general way. Each term of \re{VgD} then may be described by 
the spectral representation of its argument. To avoid the square root of 
noncommuting operators in \re{Vg}, one can use the alternative representation
\be
V=-D\W^{(\eta)}\,\frac{1}{\pi}\int_{-\infty}^{\infty}\di s\frac{1}
{(D\W^{(\eta)\dag}D\W^{(\eta)})^2+s^2}\;.
\label{Vg1}
\ee

To check the continuum limit we note that in the free case and infinite volume
with the Fourier representation $V_{\ka'\ka}=\tilde{V}(\ka)\,\delta^4\!
(\ka'-\ka)$ one gets with $\ka_{\mu}=ap_{\mu}$ at the corners of the Brillouin 
zone $\tilde{V}=-1$ and at zero
\be 
\tilde{V}\ra 1-\frac{i}{|\eta(m)|}\,\tilde{E}\Big(a
\sum_{\mu}\gamma_{\mu}p_{\mu}\Big)\quad\mbox{for}\quad a\ra0\,.
\label{TVF7}
\ee
Then imposing the requirement 
\be
\tilde{W}(-1)\ne0\,,
\label{tW-1}
\ee
because of the monotony of $E(X)\,$, doublers are suppressed for $-2r<m<0$ as 
usual.  Condition \re{tW-1} corresponds to \re{conda} since at the corners of 
the Brillouin zone the eigenvalue $-1$ of $V$ occurs.  On the lattice working 
with dimensionless quantities, the limit to be considered is $\tilde{D}/a\ra
\tilde{D}_{\rm cont}\,$. Because of $H(E(X))=X$, putting 
\be
\tilde{W}(1)=2|\eta(m)|
\label{tW+1}
\ee 
the correct result with the usual normalization of the propagator is obtained 
in this quite general way.

\section{Special Cases from Literature}\se\label{SPECA} 

\subsection{GW Relation with Constant $\rho$}

The Dirac operators satisfying the GW relation \re{GW}, which have already 
been discussed in Sec.~\ref{INT}, represent the simplest special case of
the general class. They correspond to the trivial choices $h(x)=x$ and 
$w=$ const $=2\rho\,$. This gives the spectral function
\be
f(\e^{i\vp})=-2i\rho\,\e^{i\vp/2}\;\sin\frac{\vp}{2}\,,
\ee
which in the complex plane describes the well-known circle through zero around
$\rho\,$, and which inserted into \re{specd} gives the special case \re{DN} of
\re{GAO}. According to \re{tW+1} one has to put $\rho=|m|\,$.

\subsection{General GW Relation with $[R,D]\ne0$}

The Dirac operators satisfying the general GW relation\cite{gi82}
\be
\{\ga,D\}=2D\ga RD\quad\mbox{ with }\quad[R,D]\ne0\,,
\label{gGW}
\ee
where $R\dg=R$ and $[\gamma_{\mu},R]=0\,$, do {\it not} belong to the class. 
In fact, with $\ga$-hermiticity of $D$ and $[\ga,R]=0$ from this relation one 
gets $[D,D\dg]=2D\dg[R,D]D\dg$. Thus for $[R,D]\ne0$ the Dirac operator $D$ 
is not normal, which is in contrast to what is required in \re{COM}.

It is to be noted that for such operators the analysis of the index gets 
rather subtle. For them one obtains the relation 
\be
\mbox{Tr}\big(\ga(P_j+RQ_j)\big)+\mbox{Tr}(\ga RD)=0 
\ee 
in which the projections $P_j$ need not to be orthogonal and where 
eigennilpotents $Q_j$ can occur if the dimensions of the respective  
algebraic and geometric eigenspaces differ. In a lenghty proof$\,$\cite{ke01a}
it has been shown that for the eigenvalue $\la_k=0$ of $D$ these dimensions 
are equal so that $Q_k=0$ and $P_k=P_k\dg$. Thus for zero eigenvalue the 
unwanted term with $Q_k$ disappears and $P_k$ gets orthogonal as needed.
However, the precise effect of $R$ in the term $\mbox{Tr}(\ga RD)$ remains
still to be investigated.

Dirac operators for which \re{gGW} holds occur in fixed-point QCD. There
it has been proposed to to apply the rescaling\cite{ha02}
\be
D_1=\rho\1(2R)^{\h}D(2R)^{\h}
\ee
to them, by which one can switch to the operator $D_1$ which satisfies 
\re{GW}. Thus tracing the operators $D$ with \re{gGW} back to the normal
operators $D_1\,$, one can work with $D_1$ belonging to the general class.

\subsection{Extended Fujikawa Type}

The Dirac operators of a recent extension\cite{fu02} of the Fujikawa proposal
satisfy
\be
\{\ga,D\}=\rho\1D\ga D\;\;\Phi\big((2\rho)^{-2}(\ga D)^2\big)\,,
\label{FU11}
\ee
where the operator function $\Phi$ is subject to 
\be
\Phi(X)\dg=\Phi(X) \quad\mbox{for}\quad X\dg=X\,. 
\label{FUD1}
\ee 
Using the identity $[\ga,D\dg D]=[\{\ga,\ga D\},\ga D]$ with 
\re{FU11} since $D\dg D=(\ga D)^2$ we get 
\be
[\ga,D\dg D]=0\,.
\label{FU15}
\ee 
With this we have $D\dg D=\ga D\dg D\ga$ and find 
\be
[D\dg,D]=0\,.
\label{FU14}
\ee

With \re{FU14} one sees that the Dirac operators in \re{FU11} can also be
considered as satisfying the general GW relation with $2R=\rho\1\Phi(
(2\rho)^{-2}D\dg D)\,$, where, however, in contrast to \re{gGW}, one has
$[R,D]=0\,$. 

Using the $\ga$-hermiticity of $D$ we can write \re{FU11} in the form 
\be
D+D\dg=\rho\1D\dg D\;\,\Phi\big((2\rho)^{-2}D\dg D\big)
\label{FUD2}
\ee 
and comparing with the basic relation \re{gg1}, $D+D\dg V=0\,$, it follows that
\be
V=1-\rho\1 D\,\Phi\big((2\rho)^{-2}D\dg D\big)\,.
\label{FU12}
\ee
Because of \re{FU15} and $\Phi\dg((2\rho)^{-2}D\dg D)=\Phi
((2\rho)^{-2}D\dg D)$ this is seen to be $\ga$-hermitian and using \re{FU14} 
and \re{FU11} it can be checked to be unitary. It thus turns out that the 
operators satisfying \re{FU11} are also a special case of the general class.

With \re{FU12} and \re{gg1} one gets the equation 
\be
\rho\1 D\,\Phi\big(-(2\rho)^{-2}V\1D^2\big)+V=1
\label{FU13}
\ee
for $D$ and $V$, the solution of which gives $D=F(V)\,$.

\subsection{Proposal of Fujikawa}\label{FUP}

The original proposal of Fujikawa\cite{fu00} is given by the choice
\be
\Phi(X)=X^k\,,\quad k=1,2,\ldots
\ee
of the function $\Phi$ in \re{FU11}. With this \re{FU13} becomes
\be
2(-V)^{-k}\big((2\rho)\1D\big)^{2k+1}+V=1\,,
\ee
which can be solved for $D$ giving
\be
D=2\rho\Big(\h(1-V)(-V)^k\Big)^{1/(2k+1)}\,,
\label{FU6}
\ee
where of the $(2k+1)$-th roots the one which satisfies \re{cond*o} is to be 
choosen. Thus with \re{FU6} we have indeed the form $D=F(V)$ (while in
Ref.~\refcite{fu00} only the form  $D=2\rho\ga\,(\h\ga(1-V))^{1/(2k+1)}$ 
involving also $\ga$ occurs).

The functions $f$ of the spectral representation \re{specd} for $D\,$, in the
present case are obtained from \re{FU6} as
\be
f(\e^{i\vp})=-2i\rho\,\e^{i\vp/2}\Big(\sin\frac{\vp}{2}\Big)^{1/(2k+1)}
\label{rs}
\ee
where of the $(2k+1)$-th roots the real one is to be taken. They give the
curves describing the location of the eigenvalues of $D\,$, which for $k>0$ 
arise as deformations of the circle for $k=0\,$. All of them meet the real 
axis at zero and at the value $f(-1)=2\rho\,$. 

Comparing with \re{GAi} it is seen that in the present case one has the choice 
\be 
h(x)=x^{1/(2k+1)}\,,\;\; k=1,2,\ldots
\label{GE9}
\ee
where the real one of the $(2k+1)$-th roots is to be taken. The function $w$ 
is obviously constant here, $w=(2\rho)^{2k+1}\,$, and according to \re{tW+1} 
one has to put $w=2|m|^{2k+1}\,$. 

In Ref.~\refcite{fu00} in the definition of the generalized Wilson-Dirac
operator the factors $i$ we have in \re{VgD} are not included. For 
$h(x)=x^{1/(2k+1)}$ both formulations are possible. However, in the general
case only our formulation with the strict use of hermitian functions of 
hermitian operators appears appropriate.

\section{Further Special Cases}\se\label{SPECB} 

\subsection{General $w$ with $h(x)=x$}

For $h(x)=x$ \re{GAO} specializes to
\be
D=\textstyle{\h}(\Id-V)\;W\big(\textstyle{\h}(V+V\dg)\big)\,.
\label{GAO1}
\ee
Since $E(X)=X$ for all operators of this subclass the overlap form of $V$ is
suitable.

It is instructive to consider the eigenvalues of $V$ in the free case, which
here can be calculated explicitly as those of $\tilde{V}$,
\be
\e^{i\vp}=-(\tau\pm i\sqrt{s^2})/\sqrt{\tau^2+s^2}\,,
\label{TU1}
\ee
where
\be
s^2=\sum_{\mu}\sin\ka_{\mu}^2\,,\quad \tau=m+r\sum_{\mu}(1-\cos\ka_{\mu})\,,
\quad -2r<m<0\,.
\ee
The real eigenvalues in \re{TU1} obviously occur for $\ka_{\mu}=0,\pi\,$ and 
one gets $+1$ if all $\ka_{\mu}=0$ and $-1$ at each corner of the Brillouin 
zone. Noting that
\be
\textstyle{\h}\Big(\tilde{V}(\ka)+\tilde{V}\dg(\ka)\Big)=
-\tau/\sqrt{\tau^2+s^2}=\cos\vp
\label{TU1a}
\ee
it becomes explicit that the conditions \re{tW+1} and \re{tW-1} on 
$\tilde{W}(1)$ and $\tilde{W}(-1)$ are related to the behavior of $V$ at its 
eigenvalues $+1$ and $-1\,$, repectively.

\subsection{Expansion in Powers of $V$}

In the case where $w(t_1)$ allows for an expansion in a series or a polynomial
we obtain for \re{GE4}
\be
g=s_0\;w(t_1)=s_0\;\Big(b_0/2+\sum_{\mu\ge1}b_{\nu}t_{\nu}\Big)=
2\sum_{\nu\ge0}c_{\nu}s_{\nu}\,,
\label{GE13}
\ee
which follows using \re{GE2} and \re{GE3} and relating the coefficients by 
$b_{\mu}=2\sum_{\nu\ge\mu}c_{\nu}\,$. With \re{GE13} we then have in terms of 
operators 
\be
D=\sum_{\nu\ge0}c_{\nu}(V^{-\nu}-V^{\nu+1})\,.
\label{DVt}
\ee
Condition \re{conda} now gets the form
\be
f(-1)=2\sum_{\nu\ge0}(-1)^{\nu}c_{\nu}\ne0
\label{condc}
\ee
and \re{tW+1} corresponds to
\be
\sum_{\nu\ge0}(2\nu+1)c_{\nu}=|m|\,.
\label{condb}
\ee
Obviously the special case \re{DN} of \re{DVt} arises by putting 
$c_{\nu}=\rho\,\delta_{\nu0}\,$. 

For the functions $f$ in the spectral representation of the Dirac operators 
\re{DVt} we obtain the form
\be
f(\e^{i\vp})=-2i\e^{i\vp/2}\sum_{\nu\ge0}c_{\nu}\sin(2\nu+1)\frac{\vp}{2}\,.
\label{Dla}
\ee
The eigenvalues of $D$ thus reside on a closed curve in the complex plane
which is given by a linear combination of rosette functions 
$-i\e^{i\vp/2} \sin(2\nu+1)\vp/2$ and which meets the real axis at zero 
and at the value \re{condc}. 

In case of an infinite number of terms of the expansion \re{DVt}, convergence 
properties can be studied considering the Fourier series
$\sum_{\nu\ge0}c_{\nu}\sin(2\nu+1)\frac{\vp}{2}$ in \re{Dla}.
Uniform convergence then is guaranteed by the condition
\be
\sum_{\nu\ge0}|c_{\nu}|<\infty \,.
\label{GLM}
\ee

\subsection{Nontrivial $w$ with $h(x)=x^{1/(2k+1)}$}\label{EXN}

As examples of nontrivial $w$ together with nontrivial $h$ we use
$h(x)=x^{1/(2k+1)}$ with $k=0,1,2,\ldots$ to have simple explicit forms of 
the latter. Inserting \re{SY2} into \re{specd} here gives
\be
D=\Big(\h(1-V)(-V)^k\,W\big(\textstyle{\h}(V+V\dg)\big)\Big)^{1/(2k+1)}\,,
\label{GAO2}
\ee
where of the $(2k+1)$-th roots the one which satisfies \re{cond*o} is to be 
choosen and where $W$ is subject to \re{GA1i}. Obviously \re{GAO2} generalizes
\re{FU6} replacing the constant $2\rho$ there by the function 
$W^{1/(2k+1)}\,$. The requirement \re{tW+1} now becomes 
$\tilde{W}(1)=2|m|^{2k+1}\,$.

The eigenvalues of $V$ in the free case can again be determined,
\be
\e^{i\vp}=-\Big(\tau_k\pm i\sqrt{\big(s^2\big)^{2k+1}}\Big)/
\sqrt{\tau_k^2+\big(s^2\big)
^{2k+1}}\,,
\label{TU1j}
\ee
with
\be
s^2=\sum_{\mu}\sin\ka_{\mu}^2\,,\quad \tau_k=m^{2k+1}+\Big(r\sum_{\mu}(1-
\cos\ka_{\mu})\Big)^{2k+1}\,, \quad -2r<m<0\,,
\ee
and are $+1$ if all $\ka_{\mu}=0$ and $-1$ at each corner of the Brillouin 
zone. From 
\be
\textstyle{\h}\big(\tilde{V}(\ka)+\tilde{V}\dg(\ka)\big)=
-\tau_k/\sqrt{\tau_k^2+(s^2)^{2k+1}}=\cos\vp 
\label{TU1b}
\ee
the relation of the conditions \re{tW+1} and \re{tW-1} on $\tilde{W}(1)$ and 
$\tilde{W}(-1)$ to the behavior of $V$ at its eigenvalues $+1$ and $-1\,$, 
repectively, gets again explicit.

\subsection{Polynomial Form of $\eta$(y)}

A possibility to get further concrete examples of nontrivial $h$ is to start 
from $\eta$ being a polynomial,
\be
\eta(y)=\sum_{\nu=0}^N\B_{\nu}\,y^{2\nu+1}\,,
\label{ID2e}
\ee
with real coefficients $\B_{\nu}\,$. Because of $\eta(h(x))=x$ to obtain 
$h(x)$ we have to solve the algebraic equation 
\be
\sum_{\nu=0}^N\B_{\nu}\,h^{2\nu+1}-x=0\,.
\label{ALGe}
\ee  

The solution of \re{ALGe} is trivial if only one of the coefficients 
$\B_{\nu}$ is different from zero, which leads to 
\be
h(x)=\Big(\frac{x}{\B_{\nu}}\Big)^{1/(2\nu+1)}\,,\quad\nu=0,1,2,\ldots\;.
\ee
Inserting $x=s_0w(t_1)$ with nontrivial $w$ this gives the examples of 
Sec.~\ref{EXN}.

\subsection{$h(x)$ from Cubic Equation}

In the case where only the coefficients $\B_0$ and $\B_1$ in \re{ALGe} are 
nonvanishing we have the cubic equation
\be
h^3+3ph+2q=0\,,\qquad p=\frac{\B_0}{3\B_1}\,,\quad q=-\frac{x}{2\B_1}\,.
\label{ALG3}
\ee
Requiring $\B_0\B_1>0$ it follows that $p^3+q^2>0\,$, which implies 
that one gets the real solution
\be
h=\sqrt[3]{-q+\sqrt{q^2+p^3}}+\sqrt[3]{-q-\sqrt{q^2+p^3}}\,.
\ee
We thus have in more detail
\be
h(x)=
\sqrt[3]{\frac{x}{2\B_1}+\sqrt{\Big(\frac{x}{2\B_1}\Big)^2+\Big
(\frac{\B_0}{3\B_1}\Big)^3}}\;\,\nonumber\\+\sqrt[3]{\frac{x}{2\B_1}
-\sqrt{\Big(\frac{x}{2\B_1}\Big)^2+\Big(\frac{\B_0}{3\B_1}\Big)^3}}\,,
\label{EXAe}
\ee
with the required properties for $h(x)$ of being odd and strictly monotonous. 
The related operator expression \re{GAO} for $D$ then becomes
\ba
D=\sqrt[3]{\frac{1}{4\B_1}V(V-1)W+\sqrt{\Big(\frac{1}{4\B_1}V(V-1)W\Big)^2+
\Big (\frac{\B_0}{3\B_1}\Big)^3}}\nonumber\\
+\sqrt[3]{\frac{1}{4\B_1}V(V-1)W-\sqrt{\Big(\frac{1}{4\B_1}V(V-1)W\Big)^2+
\Big (\frac{\B_0}{3\B_1}\Big)^3}}
\label{GAOe}
\ea
with $W=W\big(\textstyle{\h}(V+V\dg)\big)\,$. Condition \re{tW+1} here 
requires $\tilde{W}(1)=2|\B_0m+\B_1m^3|\,$.

\section{Index and Spectral Flows}\se\label{FLOW}

\subsection{Relation to Flows}

The spectral flows of the hermitian Wilson-Dirac operator $\Hw$ are of
fundamental importance in the overlap formalism.\cite{na93} They provide
a further description of the index of $D\,$. To make contact to the 
formulations here we note that $\Hw$ is given by $\Hw=\ga D\W$ where 
$D\W$ is $D\W^{(\eta)}$ of \re{VgD} specialized to $E(X)=X\,$. Then
according to \re{Vg} one has
\be
-\ga V=\Hw\big(\sqrt{\Hw^2}\,\big)\1=\epsilon(\Hw)\,.
\label{Vgx}
\ee
With \re{Vgx} relation \re{INN} for the index of $D$ becomes
\be
N_+(1)-N_-(1)=\h\Tr(\ga V)=-\Tr\,\epsilon(\Hw)\,.
\label{INNx}
\ee
This shows that the index of $D$ is also given by the difference of
the numbers of positive and negative eigenvalues of $\Hw\,$, which is the
view introduced in the overlap formalism.\cite{na93} It has led there to
investigations of spectral flows, i.e.~of the eigenvalues as a functions 
of the mass parameter.

We note that this view extends to the case of our general functions $E(X)\,$,
because considering the individual terms in \re{VgD} one can still confirm 
$\ga$-hermiticity of $D\W^{(\eta)}\,$. For the function with $\gamma_{\mu}$
there for this purpose one has to use the spectral representation of its 
argument.  Then with the generalized hermitian Wilson-Dirac operator 
$\Hw^{(\eta)}=\ga D\W^{(\eta)}$ instead of \re{INNx} one obtains
\be
N_+(1)-N_-(1)=\h\Tr(\ga V)=-\Tr\,\epsilon(\Hw^{(\eta)})\,.
\label{INNy}
\ee

\subsection{Differential Equation}

For the description of the spectral flows of $\Hw$ an exact diferential
equation has been derived and a complete overview of its solutions has been
given.\cite{ke99} To get this differential equation one notes that for $\Hw$ 
one has  the eigenequation
\be
\Hw \phi_l = \alpha_l \phi_l \,.
\label{egH}
\ee
Multiplying \re{egH} by $\phi_l\dg\ga$ one gets 
$\phi_l\dg\ga \Hw \phi_l = \alpha_l \phi_l\dg\ga \phi_l$ and summing this
and its hermitian conjugate one has
$\phi_l\dg \{\ga,\Hw\}\phi_l = 2\alpha_l \phi_l\dg\ga \phi_l$. From this
by inserting the explicit form of $\Hw$ one obtains
\be
 \alpha_l\, \phi_l\dg\ga \phi_l = m + g_l(m)\,,\quad
  g_l(m) = \frac{r}{2} \sum_{\mu} ||\na{\mu}\phi_l||^2  \,.
\label{gm}
\ee
For $g_l(m)$ using 
$||\na{\mu}\phi_l||\le||(\na{\mu}-\Id)\phi_l||+||\phi_l||=2$ one gets
$0\le g_l(m)\le 8r\,$.
Further, abbreviating $(\di \alpha_l)/(\di m)$ by $\dot{\alpha}_l$, we 
obtain
\be
\frac{\di (\phi_l\dg \Hw \phi_l)}{\di m}=\phi_l\dg \dot{\Hw} \phi_l +
\dot{\phi}_l\dg \Hw \phi_l+\phi_l\dg \Hw \dot{\phi}_l =
\phi_l\dg\ga \phi_l+\alpha_l \frac{\di (\phi_l\dg \phi_l)}{\di m}  
\label{dH}
\ee
which means that we have
\be
\dot{\alpha}_l=\phi_l\dg\ga \phi_l \, .
\label{pm}
\ee
Combining \re{gm} and \re{pm} we get the differential equation
\be
\dot{\alpha}_l(m) \alpha_l(m)= m + g_l(m)
\label{dif}
\ee
for the eigenvalue flows of the hermitean Wilson-Dirac operator $\Hw\,$.

\section{Index and Limit}\se\label{SUML}

\subsection{Introductory Remarks}
 
So far we have considered only relations on the finite lattice, except for 
checking the continuum limit of the propagator for free fermions in 
Sec.~\ref{REA}$\,$. In this context we have derived condition \re{tW-1} 
needed for the suppresion of doublers and condition \re{tW+1} giving the 
usual normalization of the propagator in the limit. It has turned out that 
\re{tW-1} is related to condition \re{conda} which is necessary to allow 
for a nonvanishing index. The connection of these conditions to behaviors 
at the eigenvalues $-1$ and $+1$ of $V\,$, respectively, has become 
explicit from \re{TU1a} and \re{TU1b}. 

We now address some further issues of the limit related to the index of $D\,$. 
For the sum rule \re{sum} of the index we have to make sure that it still 
holds in the continuum limit, in which the unitary space in which $D$ acts 
gets a Hilbert space (see Sec.~\ref{SFF} for precise details of the
occurring spaces). Therefore, we firstly have to define the trace $\mbox{Tr}
(\ga\Id)$, which we have used to derive the sum rule, in infinite space, too. 
Secondly we have to show that a continuous part of the spectrum of $V$ does 
not contribute. This will be done in Secs.~\ref{TRE} and \ref{COS}$\,$, 
respectively. Having thus the validity of the sum rule established also in 
the limit, the fundamental structural difference to the Atiyah-Singer 
framework it reflects is pointed out in Sec.~\ref{ATI}.

One further has to check for fermions in a background gauge field that the
r.h.s.~side of \re{INN} gives the  correct limit of the topological charge, 
\be
\h\mbox{Tr}(\ga V)\ra-\frac{1}{32\pi^2}\int\mbox{d}^4x\sum_{\mu\nu\la\tau}
\epsilon_{\mu\nu\la\tau}\mbox{ tr}\Big(F_{\mu\nu}(x)F_{\la\tau}(x)\Big)\,,
\ee
and that correspondingly for the chiral anomaly, with tr denoting the trace 
in Dirac and gauge field space only, 
\be
\h\mbox{tr}(\ga V_{nn})\frac{1}{a^4}\ra -\frac{1}{32\pi^2}\sum_{\mu\nu\la\tau}
\epsilon_{\mu\nu\la\tau}\mbox{ tr}\Big(F_{\mu\nu}(x)F_{\la\tau}(x)\Big)
\label{ANO}
\ee
holds. In case of the subclass with $H(X)=X$ for this one can rely on the 
fact that for the overlap $V$ it is safely known that one gets the correct 
results (see Ref.~\refcite{ke01} for a proof and a discussion of literature).
This is so for all $W$ because only $V$ enters relation \re{INN}. For the 
case $H(X)=X^{2k+1}\,$ a check has been presented in Ref.~\refcite{fu00a}.

\subsection{Trace Expressions}\label{TRE}

The obvious condition for defining the indicated Tr-expression is to require 
it to be the limit of the respective finite-lattice results. We thus consider 
the sequence of results of $\mbox{Tr}(\ga\Id)$ for larger and larger lattices.
Since all members of this sequence are zero, the limit is zero, too, and one 
gets
\be
\mbox{Tr}(\ga\Id)=0\,.
\label{G5}
\ee 

In practice it is desirable to have also formal descriptions leading to 
\re{G5}. An immediate possibilty is to use the factorization of the identity 
into parts related to Dirac space, gauge group space and volume ${\cal N}^4$,
and to define $\mbox{Tr}(\ga\Id):=\lim_{{\cal N}\ra\infty}\mbox{Tr}(\ga\otimes
\Id_{U}\otimes\Id_{{\cal N}^4}$). Another possibility is to introduce a
regularization suitable after taking the infinite-volume limit. For this 
purpose one can, for example, use decompositions of type $\Id=\sum_j\hat{P}_j$ 
with any orthogonal projections $\hat{P}_j$ of finite dimension and trivial in 
Dirac space (i.e.~of form $\hat{P}_j=\ga\otimes\hat{p}_j$), and define 
$\mbox{Tr}(\ga\Id): =\sum_j \mbox{Tr}(\ga\hat{P}_j)\,$. The point here 
obviously is that the summation occurs after taking the traces.

Introducing $\hat{P}_j^{\pm}=\h(1\pm\ga)\hat{P}_j$, the latter definition can 
also be written as $\mbox{Tr}(\ga\Id):=\sum_j \mbox{Tr}(\hat{P}_j^{(+)}-
\hat{P}_j^{(-)})\,$. Thus a possible alternative is obtained using functions 
$c_j(t)$ of a real parameter $t$ with $c_j(0)=1$ and $\sum_j\mbox{Tr}\big(
c_j(t)\hat{P}_j^{\pm}\big)<\infty\,$. Now with each of the sums being 
separately well-defined, because of $\mbox{Tr}\,\hat{P}_j^{(+)}=\mbox{Tr}\,
\hat{P}_j^{(-)}$ one gets $\sum_j \mbox{Tr}\big(c_j(t)(\hat{P}_j^{(+)}-
\hat{P}_j^{(-)})\big)=0\,$. Since the result is independent of $t$, letting
$t\ra0$ at the end is no problem. Therefore this provides a further definition
of $\mbox{Tr}(\ga\Id)\,$. Its principle is that of the heat-kernel 
regularization, to which we come back in Sec.~\ref{ATI}.

For the projections $P\mi{R}=\h(1+\ga)\Id$ and $P\mi{L}=\h(1-\ga)\Id$, which
project on the total right-handed and left-handed space, respectively,
$P\mi{R}-P\mi{L}=\ga\Id$ holds. Therefore, with the definition of 
$\mbox{Tr}(\ga\Id)$ in infinite space in place, which gives \re{G5}, we obtain 
\be
\mbox{Tr}(P\mi{R}-P\mi{L})=0
\label{SY}
\ee
for the difference of the respective dimensions.

\subsection{Continuous Spectrum}\label{COS}

In Hilbert space the spectra of operators can also get continuous parts. To 
show that such parts do not contribute to the sum rule \re{sum}, we note that
the derivation of this rule is based on expressions of form $\mbox{Tr}\big(
\ga\Phi(V)\big)$. To evaluate this we can use the general spectral 
representation of $V$, which is given by the operator Stieltjes integral
\be
V=\int_{-\pi}^{\pi}\e^{i\vp}\,\di E_{\vp}\,,
\ee
with which the expressions of interest are represented by
\be
\mbox{Tr}\big(\ga \Phi(V)\big)=\int_{-\pi}^{\pi}\phi(\e^{i\vp})\,\di\! 
\big(\mbox{Tr}(\ga E_{\vp})\big)\,.
\label{FV}
\ee
Its discrete part is given in terms of \re{specv}. Here it remains to consider 
the continuous part, for which we obtain 
\be
V\mi{con}=\int_{-\pi}^{\pi}\e^{i\vp}\,\di E\mii{con}{\vp}
         =\int_{0}^{\pi}\e^{i\vp}\,\di E\mii{con}{\vp}
         -\int_{0}^{\pi}\e^{-i\vp}\,\di E\mii{con}{-\vp}\,,
\label{Dcon}
\ee
where the subdivision into two integrals is possible because the projector 
function $E\mii{con}{\vp}$ is purely continuous. With $V\dg=\ga V\ga$ and 
$E\mii{con}{\vp}\dg=E\mii{con}{\vp}\,$, using \re{Dcon} one gets $\ga 
E\mii{con}{\vp}=-E\mii{con}{-\vp}\ga\,$. Further, because $E\mii{con}{\vp'}
E\mii{con}{\vp}=E\mii{con}{\vp}E\mii{con}{\vp'}=E\mii{con}{\vp}$ holds for 
$\vp\le\vp'$, it follows that 
\be
\mbox{Tr}(\ga  E\mii{con}{\vp})= 0\,. 
\ee
With this we find that the continuous part of \re{FV} indeed vanishes,
\be
\mbox{Tr}\big(\ga \Phi(V\mi{con})\big)=\int_{-\pi}^{\pi}\phi(\e^{i\vp})\,\di\! 
\big(\mbox{Tr}(\ga E\mii{con}{\vp})\big)=0\,.
\label{TT}
\ee

\subsection{Differences to the Atiyah-Singer Case}\label{ATI}
 
The sum rule for the index of the Dirac operator, which turns out to be 
important in lattice theory, is related to a basic structural difference
to the framework of the Atiyah-Singer Dirac operator.\cite{at68} Because this 
appears to be not sufficiently realized, we here point out some details. 

The definition in the Atiyah-Singer case is based on (Weyl)
operators $D^{(+)}\mi{AS}$ and $D^{(-)}\mi{AS}$ which map from the total 
right-handed space ${\cal E}^{(+)}$ to the total left-handed space 
${\cal E^{(-)}}$ and back, respectively. It is given in the combined space 
${\cal E}^{(+)}\oplus\,{\cal E^{(-)}}$ by $D\mi{AS}=\hat{D}^{(+)}\mi{AS}+
 \hat{D}^{(-)}\mi{AS}$ where $\hat{D}^{(\pm)}\mi{AS}=D^{(\pm)}\mi{AS}$ on 
${\cal E^{(\pm)}}$ and $\hat{D}^{(\pm)} \mi{AS}=0$ on ${\cal E^{(\mp)}}$. 
Because of $D^{(+)\dag}\mi{AS}=D^{(-)}\mi{AS}$ the Dirac operator $D\mi{AS}$ 
is self-adjoint and since it acts on a compact manifold its spectrum is 
discrete. Thus it is represented by $D\mi{AS}=\sum_j\la\mii{AS}{j}
P\mii{AS}{j}$ which implies $D\mi{AS}^2=\sum_j\la\mii{AS}{j}^2P\mii{AS}{j}$. 

On the other hand, one gets $D\mi{AS}^2=\hat{D}^{(+)}\mi{AS}\hat{D}^{(-)}
\mi{AS}+\hat{D}^{(-)}\mi{AS}\hat{D}^{(+)}\mi{AS}$, which can be evaluated
noting that the operators $D^{(-)}\mi{AS}D^{(+)}\mi{AS}$ and $D^{(+)}\mi{AS}
D^{(-)}\mi{AS}$ map within ${\cal E}^{(+)}$ and ${\cal E}^{(-)}$, respectively,
and are selfadjoint and nonnegative. With the eigenequation $D^{(-)}D^{(+)}
\Phi_j=\kappa_j\Phi_j$ in ${\cal E}^{(+)}$ one gets the eigenequation 
$D^{(+)}D^{(-)}(D^{(+)}\Phi_j)=\kappa_j(D^{(+)}\Phi_j)$ in ${\cal E}^{(-)}$. 
Further, from $\langle D^{(+)}\Phi_{j r'}|D^{(+)}\Phi_{j r}\rangle=
\langle\Phi_{j r'}|D^{(-)}D^{(+)}\Phi_{j r}\rangle=\kappa_j\langle\Phi_{j r'}|
\Phi_{j r}\rangle$ one sees that, except for $\kappa_j=0$, for each of the 
common eigenvalues $\kappa_j$ the eigenspaces must have the same dimension. 
Therefore, except for $\kappa_j=0$, the operators $D^{(-)}\mi{AS}D^{(+)}
\mi{AS}$ and $D^{(+)}\mi{AS}D^{(-)}\mi{AS}$ have the same spectra. Denoting 
the projections on their eigenspaces by $P\mii{AS}{j}^{(+)}$ and 
$P\mii{AS}{j}^{(-)}$, respectively, and comparing the above expressions for 
$D\mi{AS}^2$ it then follows that 
$D\mi{AS}=\sum_j\la\mii{AS}{j}(P\mii{AS}{j}^{(+)}+P\mii{AS}{j}^{(-)})$ 
and that one always has 
\be
\Nh_{+}(\la\mii{AS}{j})=\Nh_{-}(\la\mii{AS}{j})\mbox{ for }\la\mii{AS}{j}\ne0
\label{as1}
\ee
for the dimensions $\Nh_{\pm}(\la\mii{AS}{j})=\mbox{Tr}\,P\mii{AS}{j}^{(\pm)}$
of the eigenspaces. 

For $\kappa_j=\la\mii{AS}{j}=0$ one obtains
ker$\,D^{(+)}\mi{AS}=$ ker$\,D^{(-)}\mi{AS}D^{(+)}\mi{AS}$ and 
ker$\,D^{(-)}\mi{AS}=$ ker$\,D^{(+)}\mi{AS}D^{(-)}\mi{AS}$ (as is obvious 
from left to right and follows from right to left from 
$\langle\Phi|D^{(-)}\mi{AS}D^{(+)}\mi{AS}\Phi\rangle=
\langle D^{(+)}\mi{AS}\Phi|D^{(+)}\mi{AS}\Phi\rangle$). 
Thus the eigenspaces with eigenvalue zero have the dimensions
$\Nh_{\pm}(0)=$ dim~ker$\,D^{(\pm)}\mi{AS}$ and the index becomes 
\be
\Nh_+(0)-\Nh_-(0)=\mbox{dim ker }D^{(+)}\mi{AS}-\mbox{dim ker }D^{(+)\dag}
\mi{AS}\,. 
\label{as0}
\ee

It is now seen that, while according to \re{as1} pairs of subspaces with the
same dimension and opposite chirality occur for $\la\mii{AS}{j}\ne0$, with 
\re{as0} in the topologically nontrivial case the chiral subspaces for 
$\la\mii{AS}{j}=0$ contribute differently to the dimensions of the total 
right-handed and left-handed space. This is obviously different from the
situation in lattice theory, where \re{G5} (on which the sum rule is based)
reflects symmetry between right-handed and left-handed space.

To compare this in more detail with the lattice case,
we consider the projections $P\mi{R}=\sum_jP\mii{AS}{j}^{(+)}$ and 
$P\mi{L}=\sum_jP\mii{AS}{j}^{(-)}$ which here project on the total 
right-handed and left-handed space, respectively. Analogous regularizations
as in in Sec.~\ref{TRE} can be used to evaluate the trace of the difference. 
The simple prescription of taking the trace before summing gives 
$\mbox{Tr}(P\mi{R}-P\mi{L}):=\sum_j \mbox{Tr}(P\mii{AS}{j}^{(+)}-
P\mii{AS}{j}^{(-)})=\Nh_+(0)-\Nh_-(0)$. Using a function $c\,$, which 
satisfies $\sum_j\mbox{Tr} \,c(D^2t)P\mii{AS}{j}^{(\pm)}<\infty$ for $t>0$ 
and $c(0)=1\,$, with $\mbox{Tr}(P\mi{R}-P\mi{L}):=\lim_{t\ra0}\mbox{Tr}
\big(c(D^2t)(P\mi{R}-P\mi{L})\big)$ one gets $\mbox{Tr}\big(c(D^2t)(P\mi{R}-
P\mi{L})\big)=\Nh_+(0)-\Nh_-(0)$ independently of $t$, i.e.~the same result. 
For $c(x)=\exp(-x)$ this is the heat-kernel regularization frequently used
in works on the Atiyah-Singer case. Thus we here have 
\be
\mbox{Tr}(P\mi{R}-P\mi{L})=\Nh_+(0)-\Nh_-(0)\,.
\label{ASY}
\ee
The comparison of \re{ASY} and \re{SY} reveals the precise difference of the 
space structures. One shold note that \re{ASY} introduces a dependence  on the 
particular gauge field configuration, while \re{SY} does not.

\section{Cayley-Transform Type $V$}\se\label{CAY}

\subsection{Explicit Forms}

Looking for explicit forms of the unitary operator $V$ one notes that 
there are three standard constructions. They are up to constant phase 
factors given by
\begin{romanlist}[(iii)] 
\item $X(\sqrt{X\dg X}\,)\1$, normalizing an operator $X$, not
requiring particular mathematical properties of $X$ (apart from $X\dg X\ne0$), 
\item the Cayley transform $(Y-i\Id)(Y+i\Id)\1$ of
a hermitian operator $Y$, 
\item $\exp(iZ)$ with a hermitian generator $Z$. 
\end{romanlist}
In addition to unitarity one gets $\ga$-hermiticity of the constructed 
operators $V$ by requiring $X$, $iY$, $iZ$ to be $\ga$-hermitian.

Construction (i) has been used by Neuberger in the overlap Dirac 
operator\cite{ne98} and here in the general realization of $V$ by 
\re{Vg} with \re{VgD}.

With respect to (ii) we observe that it is actually the basis of a Dirac 
operator which has been introduced by Chiu\cite{ch01} in the GW case \re{GW}.
In fact, the form $D=2\rho r D_c(\Id+rD_c)\1$ in Ref.~\refcite{ch01}, with an 
appropriate antihermitian operator $D_c$ and a suitable positive constant 
$r\,$, amounts to putting
\be
V=-(Y-i\Id)(Y+i\Id)\1 \quad\mbox{ with }\quad Y=irD_c \,.
\label{CH}
\ee
This operator has turned out, due to the sum rule, not to admit a nonvanishing 
index\cite{ch01}. We have shown\cite{ke02} that the respective phenomenon 
generally occurs for Cayley-type operators on the finite lattice, while in 
the continuum limit this defect is no longer there. 

Construction (iii) has not been applied in the present context. For this one 
could think of using an operator $Z$ of the form of $Y$ mentioned above and 
tune $r$ to cover the spectrum appropriately. However, since different tunings
would be necessary for different gauge field configurations, in practice
(iii) appears not useful.

\subsection{Analysis of Cayley-Type $V$}\label{ACAY}

In our investigation\cite{ke02} of Construction (ii) above we consider the 
general choice
\be
V=-(Y-i\Id)(Y+i\Id)\1=2(\Id+Y^2)\1-\Id+i\,2Y(\Id+Y^2)\1\,.
\label{CA}
\ee
It is immediately obvious from \re{CA} that for the eigenvalue $y=0$ of $Y$ 
one gets the eigenvalue $v=1$ of $V$. 

On the finite lattice the operators act in a unitary space of finite dimension.
Therefore, requiring $Y$ to be a well-defined hermitian operator, its spectrum
consists of a finite number of real eigenvalues (which are discrete and 
finite). Introducing $s=\mbox{max}(|y\mi{min}|,|y\mi{max}|)$, where $y\mi{min}$
and $y\mi{max}$ denote the smallest and the largest eigenvalue of $Y$, 
respectively, according to \re{CA} we have
\be
\mbox{Re }v\ge\frac{2}{1+s^2}-1\;\mbox{ for all }v\,,\quad
|\mbox{Im }v|\ge \frac{2s}{1+s^2}\;\mbox{ for }\;\mbox{Re }v<0\,.
\label{CC}
\ee
We see from this that the eigenvalue $v=-1$ of $V$ cannot be reached. Thus 
on the finite lattice Construction (ii) does generally not meet the basic 
requirement needed to allow for a nonvanishing index. 

The obvious obstacle which prevents from reaching the eigenvalue $-1$ of $V$ 
is that on the finite lattice $Y$ is bounded. A related problem is that there 
the inverse Cayley transform, 
\be
Y=-i(V-\Id)(V+\Id)\1\,,
\label{BA}
\ee 
is not valid for all unitary operators but only for the subset for which the
spectrum does not extend to $-1\,$.

The crucial observation now is that the indicated restrictions no longer hold
in Hilbert space, where one gets a well-defined connection between general 
unitary operators $V$ and selfadjoint operators $Y$ which can also be 
unbounded. To recall how this comes about we start from the general spectral 
representation of unitary operators,
\be
V=\int_{-\pi}^{\pi}\e^{i\vp}\di E_{\vp}\,, 
\label{EV}
\ee 
where the projection function $E_{\vp}$ accounts for discrete as well as for 
continuous contributions. Naive insertion of \re{EV} into \re{BA} does not 
generally make sense because $-i(\e^{i\vp}-1) (\e^{i\vp}+1)\1 =\tan\!
\frac{\vp}{2}$ is not bounded, diverging for $\vp=\pm\pi$ where the value $-1$
of the spectrum of $V$ is reached. However, $Y$ is well-defined on Hilbert 
space vectors $f$ by 
\be
Yf=\lim_{{\tilde\vp}\ra\pi}\int_{-\tilde{\vp}}^{\tilde{\vp}}
\tan \!\frac{\vp}{2} \,\di (E_{\vp}f)
\label{EY}
\ee
in the sense of strong convergence. This is seen noting that with 
$f=(\Id+V)g$ one gets $\tan^2\!\frac{\vp}{2}\,\di||E_{\vp}f||^2=4\sin^2\!
\frac{\vp}{2}\,\di||E_{\vp}g||$ over which the integral from $-\pi$ to $\pi$ 
is obviously finite.

Thus with unbounded operators $Y$ in \re{EY} we indeed get unitary operators
$V$ in \re{EV} with a spectrum extending to $-1\,$, as is necessary in order
that the sum rule \re{sum} can admit a nonvanishing index. It is seen that for 
this a Hilbert space is necessary, which not only has infinite dimension but 
also includes its limit elements. Obviously the latter here is of crucial 
importance.

We thus find that on the finite lattice Cayley-type type operators generally
do not allow for a nonvanishing index, while in the continuum limit they do.
The respective problems therefore turn out to be not restricted to the 
particular realization, in which it they have been observed by 
Chiu\cite{ch01}, but to occur generally for Cayley-type constructions.

Similar remarks apply to the chiral anomaly, which on the lattice is 
given by $\h\mbox{tr}(\ga V_{nn})\,$. Not reaching the eigenvalue $-1$ of $V$ 
on the finite latice means that one remains with $P^{(\pm)}_2 \equiv0$ in 
\re{dV}, which is clearly felt by the anomaly. On the other hand, calculating
the continuum form of the anomaly, which implies that the limit has been
performed, the correct result is to be expected.

\subsection{Spaces for Fermions}\label{SFF}

For completeness we now point out in more detail how precisely the Hilbert
space considered above arises.

The fermion operators in the dimensionless formulation on the finite lattice 
act in a unitary space, in a basis of which a vector is related to a lattice
site $n$ , a Dirac index $\beta$ and a gauge field index $\alpha\,$. Taking 
the infinite-volume limit this unitary space gets of infinite dimension and, 
in order to be able to perform limits in it, it must be completed to a Hilbert
space. This space is the Hilbert space of sequences ${\bf l}_2\,$. The 
unitarily equivalent Hilbert space ${\bf L}_2(\pi;\ka)$ of functions $f(\ka)$ 
with $-\pi\le \ka_{\mu}\le\pi$ (Dirac and gauge-group indices being suppressed)
is obtained from ${\bf l}_2$ by a Fourier transformation.  Introducing the 
lattice spacing $a$ and variables $p=\ka/a$ the
space ${\bf L}_2(\pi;\ka)$ becomes ${\bf L}_2(\pi/a;p)$. By the limit $a\ra0$ 
one then gets the operators in ${\bf L}_2(\infty;p)$ from the ones in 
${\bf L}_2(\pi/a;p)$. The space ${\bf L}_2(\infty;x)$, unitarily equivalent to
${\bf L}_2(\infty;p)$, is again obtained  by a Fourier transformation. Instead 
of proceeding in the more instructive way sketched, one can also realize the 
direct way from ${\bf l}_2\,$ to ${\bf L}_2(\infty;x)$. The equivalent spaces 
${\bf L}_2(\infty;x)$ and ${\bf L}_2(\infty;p)$ are the ones one has in the 
continuum limit. 

In detail the definition of the operators of interest by the indicated limit
needs some care. Firstly, since the limit element cannot be given explicitly,
we resort to the definition by all matrix elements, i.e.~by weak operator 
convergence. Secondly, because two spaces are involved, the usual weak limit is to 
be slightly generalized. To show that this can be properly done, we introduce 
$f_a(p)=f(p)\Pi_{\mu}\Theta(\pi/a-p_{\mu})$ with $f(p)\in{\bf L}_2(\infty;p)$ 
and the operator $\hat{{\cal O}}_a\,$, requiring $\hat{{\cal O}}_a(p',p)$ of 
${\bf L}_2(\infty;p)$ to be equal to ${\cal O}_a(p',p)$ of ${\bf L}_2(\pi/a;p)$ 
for $-\pi/a\le p_{\mu}\le\pi/a$. Then $\langle f_a|\hat{{\cal O}}_a g_a\rangle$
in ${\bf L}_2(\infty;p)$ equals $\langle f_a|{\cal O}_a g_a\rangle$ in 
${\bf L}_2(\pi/a;p)$ for all finite $a$ and $\langle f_a|\hat{{\cal O}}_a 
g_a\rangle\ra\langle f|\hat{{\cal O}}g\rangle$ for $a\ra0$ defines the operator
$\hat{{\cal O}}$ in ${\bf L}_2(\infty;p)$.
 
It is now seen that the practical procedure of first calculating the desired 
matrix elements of the operators or of the functions of operators of interest 
on the infinite lattice and then performing the $a\ra0$ limit can be precisely 
formulated and justified in Hilbert space. Furthermore, one notes that the 
mapping from ${\cal O}_a$ to $\hat{{\cal O}}$ is not invertible so that the 
spectra of ${\cal O}_a$ and $\hat{{\cal O}}$ can be substantially different. 
Thus the operators $Y$ in Sec.~\ref{ACAY} can get unbounded in the limit as 
required.

\section{Chiral Gauge Theories}\se\label{CHIR}

\subsection{Basic Relations for Weyl Operators}

The chiral projection operators implicit in the overlap formalism of Narayanan 
and Neuberger\cite{na93} and used in the formulation of L\"uscher\cite{lu98}
are of form 
\be
P_{\pm}=P_{\pm}\dg=\h(1\pm\ga)\Id\,,\quad \tilde{P}_{\pm}=\tilde{P}_{\pm}\dg=
\h(1\pm \ga V)\Id\,,
\label{PR} 
\ee
with $V\dg=V^{-1}=\ga V \ga\,$. Obviously only $\ga$ and $V$ are involved in
\re{PR} so that we can start with it only requiring \re{gg} and \re{ga5}
for $D\,$. From condition \re{gg} we get the identity $D=\h(D-\ga D\ga V)$ 
and inserting $\ga=P_+-P_-$ and $\ga V=\tilde{P}_+ -\tilde{P}_-$ into it 
we obtain
\be
D=P_+D\tilde{P}_-+P_-D\tilde{P}_+\,.
\label{DP}
\ee
With this we have the relations 
\be
P_{\pm}D\tilde{P}_{\mp}= D\tilde{P}_{\mp}= P_{\pm}D \,,
\label{PD}
\ee 
which generalize the expressions for the Weyl operators in terms of the Dirac 
operator familiar in continuum theory. 

With respect to possible forms of \re{PD} one should be aware of the fact that
the relations 
\be
P_{\pm}\ga=\pm P_{\pm}\,,\quad \ga V\tilde{P}_{\mp}=\mp\tilde{P}_{\mp}\,, 
\label{PgV}
\ee
allow to absorb parts of $D\,$. In the special case of the Dirac operator 
\re{DN}, with \re{PgV} one gets $P_+\rho(1-V)\tilde{P}_-=2\rho P_+
\tilde{P}_-\, $, which relates the different forms of the chiral determinant 
in Ref.~\refcite{lu98} and in Ref.~\refcite{na93}.
Considering the general class of operators $D$ here, we have to observe
that \re{PD} is the generally valid form and that modifications by \re{PgV} 
depend on the particular choice of $D\,$.

\subsection{Degrees of Freedom}

For the numbers of the degrees of freedom $\Tr\,P_+$ and $\Tr\,\tilde{P}_-$ 
of the Weyl fermions in $P_+D\tilde{P}_-$ one gets from \re{PR}
\be
\Tr\,P_+-\Tr\,\tilde{P}_-=\h\Tr(\ga V),
\label{INNw}
\ee
which agrees with the result \re{INN} for the index of the Dirac operators 
$D$ of the general class. 

These degrees of freedom are exhibited in more detailed form representing the 
projections by
\be
P_+=\sum_ju_ju_j\dg\,,\quad\tilde{P}_-=\sum_k\tilde{u}_k\tilde{u}_k\dg\,,\quad
u_i\dg u_j=\delta_{ij}\,,\quad\tilde{u}_k\dg\tilde{u}_l=\delta_{kl}\,.
\label{uu}
\ee 
Clearly the choice of the bases here is not unique, however, different ones 
of them must represent the same projection, respectively, and thus are 
related by unitary transformations,
\be
v_j=\sum_lu_lS_{lj}\,,\quad\tilde{v}_k=\tilde{u}_l\tilde{S}_{lk}\,. 
\label{BTR0}
\ee 
According to \re{uu} the basis vectors are normalized vectors satisfying 
the eigenequations
\be
P_+u_j=u_j\,,\quad\tilde{P}_-\tilde{u}_k=\tilde{u}_k\,,
\label{EIG}
\ee 
or equivalently
\be
\ga\, u_j=u_j\,,\quad\ga V\,\tilde{u}_k=-\tilde{u}_k\,.
\label{EIG1}
\ee 
This shows that, apart from $\ga\,$, only the unitary operator $V$ is 
involved in the determination of $\tilde{u}_k$ for the general class 
of operators considered here.

\subsection{Correlation Functions}

Associating Grassmann variables $\bar{\chi_j}$ and $\chi_k$ to the degrees
of freedom, the fermion field variables get
\be
\bar{\psi}=\sum_j\bar{\chi}_ju_j\dg,\quad \psi=\sum_k\tilde{u}_k\chi_k\,.
\ee
The fermion action then is given by
\be
S_\f=\bar{\psi}D\psi=\sum_{j,k}\bar{\chi}_jM_{jk}\chi_k\,,
\ee 
where one has for the matrix $M$
\be
M_{jk}=u_j\dg D\tilde{u}_k\,.
\label{MM0}
\ee 

Considering fermionic correlation functions $\langle\psi_{\sigma({\sy 1})}
\bar{\psi}_{\sigma(r_1)}\ldots \psi_{\sigma({\sy f})}\bar{\psi}_{\sigma(r_f)}
\rangle_\f$ with equal numbers of fields $\psi$ and $\bar{\psi}$ ($\sigma$ 
standing for the combination $(n,\alpha,\beta)$ with $n$, $\beta$ and 
$\alpha$ being related to position space, Dirac space and gauge-group space, 
respectively) one obtains
\ba
\langle\psi_{\sigma({\sy 1})}\bar{\psi}_{\sigma(r_1)}\ldots
\psi_{\sigma({\sy f})}\bar{\psi}_{\sigma(r_f)}\rangle_\f=\hspace*{65mm}
\nonumber\\ \int\prod_l(\di\bar{\chi_l}\di\chi_l)\;\exp(-S_\f)\;
\psi_{\sigma({\sy 1})}\bar{\psi}_{\sigma(r_1)}\ldots
\psi_{\sigma({\sy f})}\bar{\psi}_{\sigma(r_f)}=\hspace*{30mm} \nonumber\\
\sum_{s_1,\ldots,s_f}
\epsilon_{s_1s_2\ldots s_f}\; (\tilde{P}_-D\1 P_+)_{\sigma({s_1})\sigma(r_1)} 
\ldots(\tilde{P}_-D\1 P_+)_{\sigma({s_f})\sigma(r_f)}\;\;\det M\;\qquad
\label{COR}
\ea
for gauge-field configurations with $\Tr\,\tilde{P}_-=\Tr\,P_+\,$, while one
gets 
\be
\langle\psi_{\sigma({\sy 1})}\bar{\psi}_{\sigma(r_1)}\ldots
\psi_{\sigma({\sy f})}\bar{\psi}_{\sigma(r_f)}\rangle_\f=0
\ee
for ones with $\Tr\,\tilde{P}_-\ne\Tr\,P_+\,$. Then including the gauge-field
integrations the full correlation functions become
\be
\frac{\int[\di U]\exp(-S_{\g})\langle\psi_{\sigma({\sy 1})}
\bar{\psi}_{\sigma(r_1)}\ldots \psi_{\sigma({\sy f})}\bar{\psi}_{\sigma(r_f)}
\rangle_\f} {\int[\di U]\exp(-S_{\g})\langle1\rangle\f}\;.
\label{CORR}
\ee

We note that the operator product $\tilde{P}_-D\1 P_+$ in \re{COR} in 
the presence of zero modes of $D$ is not yet defined. Replacing 
$D\1$ by $(D-\zeta)\1$ there and letting the parameter $\zeta$ go to 
zero after the evaluation we get a well-defined finite result. 
This follows since with \re{PPg} and \re{specv} we obtain 
$P_+P_1\tilde{P}_-=0$ for the projector $P_1=P_1^{(+)}+P_1^{(-)}$ 
on the space of zero modes of $D\,$.

\subsection{Remarks on the Chiral Determinant}

To study gauge-invariance properties of the chiral determinant $\det M$ 
in \re{COR}, in Ref.~\refcite{lu98} the variation $\delta\ln\det M =
\Tr (M\1\delta M)$ has been considered. For this expression using \re{uu} 
and \re{PD} 
\be
\Tr(M\1\delta M)=\Tr(\tilde{P}_-D\1\,\delta D)+
\sum_k\tilde{u}_k\dg\delta\tilde{u}_k
\label{GA}
\ee
is obtained, which with $\delta D= [\G,D]\,$, where $\G$ is the generator 
of the gauge transformation, becomes
\be
\Tr(M\1\delta M)=\h\Tr(\ga \G V)+
\sum_k\tilde{u}_k\dg\delta\tilde{u}_k\,.
\label{GA1} 
\ee
We note, however, that this only holds in the absence of zero modes of $D\,$,
because only then 
\be
(M\1)_{kl}=\tilde{u}_k\dg D\1 u_l
\label{MM}
\ee 
is properly defined. To get the appropriate definition in the presence of 
zero modes, too, we replace $D\1$ in \re{MM} by $(D-\zeta\Id)\1$. Then in 
the evaluation using 
\be
D\,(D-\zeta\Id)\1\ra\Id-P_1\,,\quad -\zeta\,(D-\zeta\Id)\1\ra P_1\quad
\mbox{for}\quad\zeta\ra0
\ee
we obtain the further terms
\be
-\Tr(\ga \G P_1)+\sum_k\tilde{u}_k\dg\ga P_1\delta\tilde{u}_k
\label{GAz}
\ee
which are to be added on the r.h.s.~of \re{GA1}. The projector 
$P_1=P_1^{(+)}+P_1^{(-)}$ on the space of zero modes of $D$ according 
to \re{specv} and \re{specd} may be expressed in terms of $V$ by
\be
P_1=\lim_{\zeta\ra0}-\zeta\,(V-\Id-\zeta\Id)\1\,.
\ee 
It is seen now that, apart from $\ga$ and $\G\,$, also only $V$ enters 
$\Tr (M\1\delta M)$ for the general class of operators considered.

With $\G_{n'n}=i\delta^4_{n',n}\sum_{\ell}\omega_n^{\ell}T^{\ell}\,$,
where $T^{\ell}$ are hermitian generators and the $\omega_n^{\ell}$ real,
the first term on the r.h.s.~of \re{GA1} can be written in more detail as
\be
\h\Tr(\ga \G V)=i\sum_{n,\ell}\omega_n^{\ell}\,\h\mbox{tr}(\ga T^{\ell}
V_{nn})\,.
\label{GAd}
\ee
The trace expression $\mbox{tr}(\ga T^{\ell}V_{nn})$ here differs from 
that in \re{ANO} only by the insertion of the factor $T^{\ell}\,$. Since 
the inclusion of such factor in the derivation of \re{ANO} for the chiral 
anomaly is straightforward, it becomes obvious that in the continuum limit 
one gets for the gauge anomaly
\be
\h\mbox{ tr}(\ga T^{\ell}V_{nn})\frac{1}{a^4}\ra -\frac{1}{32\pi^2}
\sum_{\mu\nu\la\tau}\epsilon_{\mu\nu\la\tau}\mbox{ tr}\big(T^{\ell}
F_{\mu\nu}(x)F_{\la\tau}(x)\big)\,,
\label{CONDA}
\ee
which is thus seen to derive generally from $V$ (and not from $D$).

\section*{Acknowledgement}

I wish to thank Michael M\"uller-Preussker and his group for their kind 
hospitality.

\end{document}